# Criteria-Based Evaluation Framework for Service-Oriented Methodologies

Mehdi Fahmideh Gholami
Computer Engineering Department, Faculty of Engineering, Science & Research Branch of Islamic Azad University, Tehran, Iran
m.fahmideh@srbiau.ac.ir

Jafar Habibi
Computer Engineering Department, Sharif University of Technology, Tehran, Iran
jhabibi@sharif.edu

Fereidoon Shams
Computer Engineering Department, Faculty of Electrical & Computer Engineering, Shahid Beheshti University, Tehran, Iran,
f_shams@sbu.ac.ir

Sedigheh Khoshnevis
Islamic Azad University, Shahr-e-Qods-Branch, Iran
s_khoshnevis@sbu.ac.ir

***Abstract*—Service-Oriented Software Engineering is based on concepts and principles for constructing complex enterprise systems in which services as building block of the system, are distributed in large networks. The main goal of the service-oriented methodologies is to define a process for development and maintenance of service-based systems. Most of the Service-Oriented methodologies are not mature enough compared with traditional software development methodologies such as Object-Oriented or Component-Based. Hence, defining an evaluation framework will be useful for comparing methodologies for identifying their strengths and weaknesses, defining new methodologies or extending existing Service-Oriented methodologies. At the time being, there is no complete evaluation framework for evaluating Service-Oriented methodologies. The principal objective of this paper is to introduce a comprehensive evaluation framework for evaluating Service-Oriented methodologies. This evaluation tool is appropriate for methodology engineers to develop new methodologies, as well as project managers to select an appropriate methodology at a specific project.**

***Keywords- Software Development Methodologies; Service-Oriented Development Methodologies; Methodology Evaluation Framework; Evaluation Criteria***

## I. INTRODUCTION AND MOTIVATION

Service-Oriented Computing is a new computing paradigm in which, services as highly-reusable building blocks are used for rapid, low-cost development of distributed applications in heterogeneous and loosely coupled environments [1]. Services shape loosely-coupled large scale distributed systems in which, services interact through their interfaces to realize business processes. Today, many service oriented methodologies have been designed for developing service-oriented (SO) systems by different companies and practitioners. SO methodologies aim at translating enterprise business processes to a set of services by defining appropriate activities, techniques, guidelines, roles and responsibilities. However, they are quite new and immature in a way that it is too difficult to select one appropriate methodology among them. Therefore, it could be said this is a time of "Methodologies War" for service - oriented engineering context [2]. IBM SOAD [3], IBM SOMA 2008 [4], CBDI-SAE [5], SOUP [6], MASOM [7], SOA RQ [8], BPMN to BPEL [9] and Papazoglou [10] are the most known SO methodologies that are proposed so far. In general, a methodology is defined by two aspects [11,12]:

***1. Development Process***: Containing techniques (for cost estimation or requirements engineering), guidelines, activities, roles and their responsibilities, verification and validation mechanisms, quality assurance for produced artifacts, metrics, coding standards and tools. Each methodology has a set of umbrella activities for controlling and managing the development process (e.g. risk management or project monitoring).

***2. Modeling Language***: Is used to represent artifacts which are produced in the development process phases.

Almost every SO methodology is defined based on common traditional methodologies such as object-oriented and component based methodologies. Thus, the SO methodologies can be considered as an evolution of the Component-Based and Object-Oriented methodologies [1,32]. SO methodologies are not basically very different from traditional software development processes, however, there are new considerations that should be taken into account in terms of new features and activities that are specific to SO methodologies. These can be represented via an evaluation framework for SO methodologies. An evaluation framework is an appropriate tool to discover weaknesses, strengths, similarities and differences of methodologies. It may be used by two groups of people in a software development organization: project managers, who select the most appropriate methodology for developing their service-oriented systems, and Methodology engineers, who extend or adapt a methodology for special situations. Moreover, an evaluation framework can assist methodology engineers to select suitable method-fragments of methodologies to create a new specific SO methodology for a specific purpose.

Our studies show that there are few evaluation frameworks to meet the above needs. For this purpose, we define a quite comprehensive evaluation framework based on a set of criteria to evaluate SO methodologies. The proposed framework criteria strive to cover most aspects of SO methodologies.

The rest of the paper is structured as follows: Section 2 overviews the related work. In section 3, we will define the structure of the proposed methodology evaluation framework. Section 4 contains detail description of evaluation criteria. In section 5 we evaluate MSOAM methodology by our proposed evaluation framework. Finally section 6, is related to opportunities for future researches.

## II. Background and Related Work

The set of evaluation criteria for evaluating a methodology should focus on the specific approach, for which, the methodology has emerged, or by which, it is influenced. For example object-oriented methodologies are defined based on the concepts such as object, attribute, operation, encapsulation, information hiding and inheritance. Agile methodologies are established base on agile manifesto [13]. Therefore, these methodologies should be evaluated by criteria which are focused on the mentioned concepts. In a similar way, SO methodologies have emerged from service-oriented architecture (SOA) concepts and thus their evaluation criteria should comply with SOA concepts.

Unlike SO methodologies, there are many researches on evaluating and analyzing methodologies based on other types of approaches: for example there are evaluation frameworks for the Object-Oriented [14,15,16,17], Agile [18,19] and Agent-Oriented methodologies [20,21]. Although SO methodologies are becoming increasingly important, there is only a little work on defining evaluation frameworks in the context of SO methodologies. In [2], most prominent SO methodologies are evaluated by a set of proposed criteria. These criteria are categorized into two categories: those related to the development process and those related to some of SO characteristics. However, the proposed criteria are not enough to cover all service orientation aspects, because some important special features of service-oriented engineering are not addressed as discussed on [1,10]. Moreover, one important deficiency of their evaluation framework is that it does not adequately address criteria that are related to the modeling language aspect of the methodology. In [1], some challenges in the service oriented software engineering are reviewed including business modeling techniques, service composition, gap analysis techniques and design principles. As a result, a lack of a comprehensive and adequate evaluation framework for SO methodologies is felt. In this paper, a pioneering research in the area of evaluation framework for SO methodologies is presented.

## III. Proposed Evaluation Framework

We provide a brief overview of the structure of our evaluation framework in this section. The following are the objectives of our proposed evaluation framework:

- It must act as a tool for project managers to select the most appropriate SO methodology in the context of development, based on a set of predefined requirements. The framework must facilitate methodology selection via clarifying similarities, differences, strengths and weaknesses of SO methodologies.
- It must help methodology engineers to extend or adapt SO methodologies in order to improve them or even construct a project-specific SO methodology by selecting appropriate methodology fragments into new different methodologies. All criteria in the evaluation framework can be considered as method fragments of a super methodology.

The first step to evaluation is to define criteria and then structure them in an appropriate evaluation framework. For this purpose, we have used DESMET. It is an evaluating approach, developed in 1996 in a collaborative project between academia and industry, and helps a methods/tools evaluator to carry out an evaluation for identifying the best option. According to DESMET there are nine methods/techniques for developing an evaluation framework [22]:

1. **Feature Analysis (Qualitative Analysis):** It is a qualitative method. Feature Analysis is based on an evaluation criteria set, that are based on the prominent features of tools, processes or products. This technique defines how to plan a feature-based evaluation and how to analyze the results of an evaluation. There are four techniques in feature analysis as listed below:
    a. **Formal Experiment:** Results of applying tools/processes/products can be analyzed using standard statistical techniques.
    b. **Case Study:** The tool/process/product is used in a real world project and effects of applying it are evaluated.
    c. **Survey**: Practitioners and researchers' opinions about a subject are used.
    d. **Screening Mode**: The evaluation is performed by researcher based on available resources and documents.
2. **Quantitative Evaluation:** This technique is based on measurable properties to select appropriate tools/processes/products.
3. **Qualitative Effects Analysis:** It uses a combination of feature analysis and quantitative techniques.
4. **Benchmarking:** This technique includes designing the benchmark tests and running them on each tool/process/product.

Our evaluation framework is developed using feature analysis (specifically screening mode). For achieving suitable and precise evaluation criteria, we used meta-criteria for eliminating overlapping, redundancy and inconsistency between criteria. Meta-criteria define essential features that an appropriate evaluation framework should deal with. In [14,24] some meta-criteria are defined for achieving appropriate evaluation criteria such as precision, simplicity, consistency, and minimum overlapping between the criteria.

Although our proposed evaluation framework is qualitative, it can be applied for comparative analysis in a quantitative manner. For this purpose, the criteria set can be formulated as mentioned in [19,25] or other techniques of DESMET such as case study or survey can be used. However, precision of each technique depends on the available resources and execution conditions of evaluation. In this paper, we use feature analysis to study the applicability of our framework by performing it on the MSOAM methodology which is proposed by Erl [7]. Although the result of this evaluation is fully subjective, other techniques can be used to achieve higher precision in a comparative analysis.

We have organized our proposed framework in five aspects (figure 1). This structure has leveraged the methodology definition, stating that a methodology has two main aspects: development process and modeling language. However, we extend the definition by adding new aspects, specific to SO methodologies. The new aspects are: Service-Oriented Activities, Service-Oriented Umbrella Activities and Supportive Features. In order to cover process and modeling language criteria we use previous researches that were done in Object Oriented [15,16 ,17], Agile [18,19] and Agent-Oriented methodologies [20,21,25,26], because these criteria are general enough to be applicable for comparative evaluation of all types of methodologies at any approach. However, there are some specific aspects in SO methodologies that are not satisfied by the general criteria. Hence, as the main feature of our research, we have defined an appropriate criteria set to address these aspects. As shown in figure 1, Service-Oriented Activities are specific criteria set that development process in SO methodologies should include. In addition, in SO methodologies there are more umbrella activities compared with common traditional methodologies. Consequently, Service- Oriented Umbrella Activities aspect has been developed for this purpose. Finally, Supportive Features aspect refers to criteria that associate with SOA concepts and service-based systems. In order to develop all criteria in these new three aspects, according to prescribed steps in DESMET, we had a precise study on existing SO methodologies, relevant resources and evidences, SOA challenges [1] and prominent features of existing SO methodologies [2]. Then we defined initial criteria and completed them through gradational manner by eliminating overlaps, redundancies and inconsistencies among criteria. Criteria definition process proceeded by categorizing them in Service- Oriented Activities, Service-Oriented Umbrella Activities and Supportive Features. Further, we described all criteria as organized into framework's aspects.

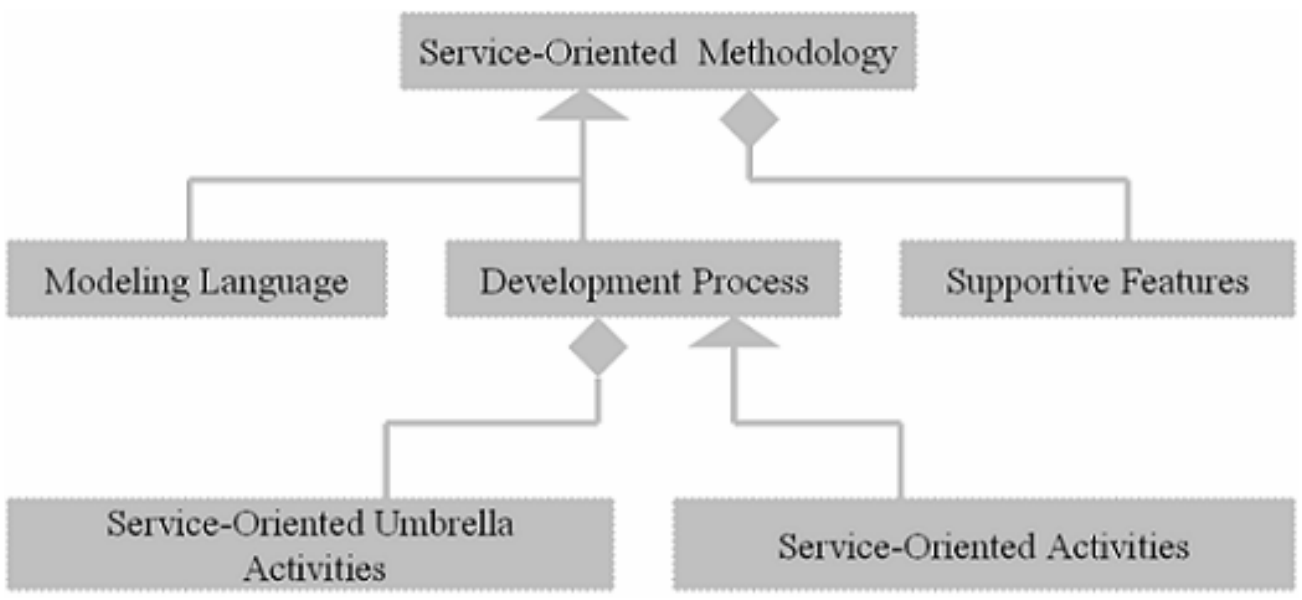

Figure 1. Class-Based Structure of Evaluation Framework

## IV. Evaluation Criteria

In this section, we describe details of our evaluation framework criteria for SO methodologies.

### A. *Development Process*

In [15,16,17,18,19,20,21,25], sets of generic criteria are defined that are applicable for evaluating the development process part of any type of methodology. Having had an overall study on them, we leveraged and used these criteria as listed below:

***1- Clarity and Consistency of Definition:*** Methodology should have precise and clear definition to be usable in a software development organization. Activities, roles and guidelines should be defined explicitly and unambiguously so that development team has no problem for using it. It is an urgent need that methodology documents and textbooks are available and that the methodology is supported adequately by its creator in order to be usable in practice, otherwise it will be inconvenient to use. Moreover, the number of success reports of applying the methodology in the industry should be considered. For example MSOAM methodology has a clear and step-by-step definition and is fully documented, that has made it easy to use. In contrast, IBM SOAD's 2004 process has a poor definition for phases and activities.

***2- Coverage of the Generic Development Lifecycle Activities:*** Methodology should have coverage for the whole generic SO life cycle including service analysis and modeling, service design, service realization, service implementation and test, service deployment and service monitoring in general. From this viewpoint, SOMA 2008 and CBDI-SAE cover about SO generic life cycle according to the available documented resources.

***3- Support Umbrella Activities:*** These activities typically include project management, risk management and quality assurance management.

***4- Smoothness and Seamlessness Transition between Phases, Stages and Activities:*** Transition between different phases and stages should be smooth and successive. Important factors by which transition smoothness is provided for a methodology are: continuous refinement of a specific set of models (e.g. Catalysis), Fractal modeling (e.g. SOMA 2008), and iterative-incremental nature of development process and the short cycles (Agile methodologies). In addition, a seamless methodology should try to avoid a huge gap and complex transformation between stages. For example many agile methodologies cannot be seamless since, after the requirement analysis phase, a quick transition is made to the implementation phase which is started immediately without analyzing and modeling the requirements.

***5- Requirements as a basis (Functional and Non-Functional):*** Methodology should be started with requirements elicitation. All activities should be based on the requirements.

***6- Tangibility of Artifacts, and Traceability to Requirements:*** Understandability of the artifacts produced by methodology and concrete relationships between them for users and development team involved in development process is referred to as tangibility [17]. A methodology that provides a variety of models, such as use-case diagrams, prototypes and graphical models is more understandable for users. In contrast, a methodology that produces complex artifacts is not useful. Artifacts should complement each other throughout the development process. Traceability of requirements leads to profitability of models. Artifacts should be traceable to the main requirements. Each artifact should show part of stakeholders' requirements.

***7- Manageability of Complexity:*** The techniques or guidelines that are defined by the methodology for breaking a large activity into simple and light sub-activities are referred to as complexity management. Methodology should provide a solution or managing complex activities such as architecture design or requirements elicitation. In MSOAM, for example, there is a step-by-step method for extracting and designing services from business processes.

***8- Extensibility, Configurability, Flexibility and Scalability:*** Extensibility of the methodology is defined as mechanisms that are explicitly defined by the methodology to add new method fragments to its development process according to the project at hand. For example, IBM SOMA's solution templates have predefined extension points for customizing its process to make it fit to a SO specific project. Configurability is a desirable feature by which, before starting its process, methodology can be configured according to the project at hand and the software development organization. CBDI-SAE and SOMA 2008 are examples of configurable methodologies. Flexibility is configurability at the methodology process runtime. If the methodology can be tuned regarding the experiences gained during the development, it is said to be flexible. Utilizing an iterative and incremental process, continuous review sessions by the development team, flexible project plan and feedback-based revisions are useful techniques that help a methodology to be flexible. Finally, scalability indicates the ability of the methodology to handle applications with different sizes and criticalities.

***9- Practicability and Practicality:*** Practicability refers to employability of the methodology in an efficient and effective manner. Using a methodology with heavy process is not practicable unless it is configured before running its process. For example, RUP, as an object-oriented methodology, is not practicable per se. It should be configured according to the project situation. Practicality of a methodology depends on its feasibility. Need to expert people, need to specific tools, learning curve, degree of complexity of development process and modeling language, and pragmatic techniques are more important issues that largely affect methodology practicality. For example, almost all of the agile methodologies such as XP, DSDM and ASD are heavily dependent on tools for having a practical agile process.

***10- Application Scope:*** Application scope relates to the intended usage context. For example in SO, there are wide spectrums of applications such as SO tele-communication, SO distributed real time systems, e-health and e-government. Each type of these systems has its own issues and challenges. A methodology should be evaluated for its suitability in a particular SO domain.

***11- Evolutionary vs. Revolutionary:*** This criteria, evaluates compatibility of the methodology with the traditional methodologies. A methodology can be defined based upon existing methodologies or engineered from scratch by novel ideas. Especially SOUP's method fragments are selected from RUP and XP methodologies. In contrast, MSOAM has been designed from scratch by Erl [7].

***12- Language or Technology:*** A methodology could be designed based on specific concepts of a programming language or technology. Therefore, software development environment and the required expertise are important factors in methodology applicability.

### *B. Modeling Language*

In this section, we will review the criteria set for evaluating the modeling language part of the methodology. These criteria are taken from [15,16,17,18,19,20].

***1- Support for Different Model Views:*** A modeling language should be able to represent elements and different aspects of the system such as behavior, structure and function.

***2- Analyzability*:** A desirable feature of a modeling language is its degree of support from formalism. It provides the modeling language with the capability to provide a symbolic execution of the system. This will help to analyze specific aspects of system behavior such as performance or availability at runtime. Formalism helps to a better automatic software construction. A tangible example in this context is OCL, which is defined within UML.

***3- Providing Techniques for Tackling Model Inconsistency and Managing Model Complexity*:** A modeling language should be able to manage complexity and inconsistency in models. For example in UML, package is a suitable mechanism for managing model complexity by bringing related classes together into a coherent package.

***4- Preciseness:*** This criterion refers to unambiguity of the modeling language. It mitigates the misinterpretation of the models for modelers.

***5- Simplicity to Learn and Use:*** A modeling language should be simple to learn and use.

### *C. Service-Oriented Activities*

This category of criteria focuses on the specific context activities (such as tasks, techniques or guidelines) that should be included in an appropriate SO based development process. We have defined these criteria based upon previous researches [1,2], service-oriented software engineering literature, SO methodologies challenges, SOA concepts and good features of existing service oriented methodologies. In our opinion, it is essential for any SO methodology to regard eight criteria as listed below:

***1- Business Modeling:*** Generally, services are identified and realized based on business processes. Business processes should be modeled and optimized in order to be mapped appropriately to the services that satisfy business goals. Better business processes alignment has direct affect on the enterprise goals. Therefore, we have defined criteria for evaluating prescribed activities, techniques and guidelines in the methodology for specifying, decomposing, optimizing and aligning the enterprise business processes with business goals.

***2- Service-Oriented Analysis & Design (SOAD)*:** Indeed, service analysis and design is the heart of any SO methodology. In SOAD, business processes are mapped to software services. In many cases, SOAD is called as methodology by itself. However, SOAD is not everything for a SO-based development process. Methodology should define a precise step-by-step process, techniques and guidelines for identifying services from business processes. Furthermore, a methodology should take into account defining techniques to enable reusing significant services from legacy systems, to be used in new business processes, leading to an increase of reusability and a reduction of development cost and time. Service discovery techniques should not be missed in the development process. When well-defined formal description is provided for the services, they can be published to a repository. Search techniques would facilitate the search for services within local or remote, centralized or distributed repositories. Tool support for automating discovery is important. The basic key elements for supporting SOAD are: *"Identifying right services at each layer", "Optimizing service granularity", "Designing fine-grained service operations and interfaces", "Service composition for orchestration or choreography", "Identifying useful existing services from legacy systems"*. In addition, an important feature of a service is adherence to SOA principles [7], such as *service reusability, statelessness of service*, etc., that are all required throughout SOAD.

***3- Service Quality Attributes*:** Services can be naturally used by every external enterprise and software development organization. Quality attributes of services such as security, availability, reliability, reusability and performance directly affect the usability of the services by their consumers. Hence, service quality should not be neglected in the architectural design. A methodology should address activities, guidelines and quality metrics for satisfying service quality at design time. Applying service-oriented patterns can be useful in this context to gain qualified services at design time.

***4- Service Provisioning and Consuming*:** Service provisioning is deploying, making available and supporting the service during the use for its consumers. It involves service provider in service publishing, certificating, enrolling, auditing, metering, billing and managing operation of services when consumers are subscribed to services. In such case, methodology should have activities and guidelines for service provisioning and consuming aspects.

***5- Service Testing*:** Test methods in service-orientation are different from traditional ones. Since services are distributed and loosely-coupled and because asynchronous messaging is used in SOA, testing can be significantly more complex in service-based systems compared with traditional testing techniques that we have used so far. There is no full control on all of the distributed services that have realized a service oriented system. It is possible that a composite service consists of some fine-grained services that have been developed and deployed independently from each other in a large network. In testing a composite service, availability of all services at testing time has an important effect on the test success. Unfortunately, a composite service, in which services can be added, removed and modified dynamically during use of the service, is more difficult to test. This criterion evaluates the methodology support for test case writing techniques and guidelines in different levels such as *component, service, composition, orchestration* and *choreography, governance* and testing the quality of services such as *security*, *performance* and *availability*.

***6- Service Versioning and Evolution*:** Each service has its own life cycle: identification, specification, realization, publishing, using and finally retirement. Many service consumers use services in a common manner. A tiny change in the service interfaces would tend to cause ripple effects on distributed business processes chain in a way that all requests based on the old interfaces would fail. A versioning management system is needed due to the adherence to predefined expected contracts. Methodology should define needed roles and activities to help the conduction of service versioning management, because occasional changes that may occur in service-interfaces can cause changes in service implementation.

***7- Adaptable with Legacy Systems*:** Legacy systems are known as existing software assets that have been developed in the past. Legacy systems generally consist of business logic that could not be neglected to be incorporated with new SO systems. However, they suffer from undocumented, outdated technology, monolithic architecture and inflexibility. To satisfy this criterion, a methodology should include prescribed techniques for *gap analysis between legacy and target SO systems*, *development of the most appropriate migration plan from legacy systems to SO environments*, *required improvements that must be made to the legacy systems to accomplish the migration*, and *databases migration*.

***8- Cost Estimation:*** There are factors that complicate estimating an SO project scope, cost and time, system boundary identification, social, cultural and organizational issues, business processes complexity, services complexity and the organization maturity level (experience in undertaking SOA). There is a need to more research to enhance traditional estimation methods for SO project. This criterion checks whether a methodology has techniques for SO-based cost estimation.

### *D. Service-Oriented Umbrella Activities*

This section deals with the specific SO based umbrella activities of a SO methodology. Our framework suggests these criteria as described below:

***1- Service Level Agreement (SLA) Monitoring*:** An SLA is a compliance contract between an IT service provider and the consumer. For example, "percentage of service availability in a month", "time to recovery from failure" and "expected service response time" are contracted in an SLA. An SLA process involves gathering, reporting, behavioral monitoring and detecting violations of the services from SLA's contracts for nonfunctional quality attributes after the services are deployed in the network environment. Adherence of a methodology to SLA is measured by checking if there are appropriate activities, techniques, guidelines, measurable metrics and supportive roles for monitoring service quality attributes, such as performance, security and availability to be maintained in a certain level at runtime.

***2- Support of Governance***: Governance is an important foundation for successful construction of SO systems. Governance is what ensures stakeholders that the right services are developed complying with diverse business goals set by different business units. Governance is an umbrella activity in the entire system development. To support governance in a methodology, we should define activities for establishing communication between development teams and stakeholders, and controlling mechanisms for enabling people to carry out their responsibilities and monitoring the execution of policies. Governance should be present in the entire development process to keep services aligned with business objectives and stakeholders' needs.

***3- People Management***: "Fear of the unknown in people is the greatest contributor of resistance to change and is made as project-killer" [27]. People in enterprises are afraid of changes and this results in human factor infeasibility; therefore, there is a tendency for project failure. Suitableness of methodology for applying in enterprise scale is ascertained with activities for addressing people management, avoiding human factor infeasibility and providing adequate training.

***4- Distributed Software Development (DSD) Techniques:*** Software services can be developed by fully distributed teams and each team has members at multiple locations. Methodology should define activities and techniques for project management to track status and progress of project across distributed teams. We can examine DSD support by checking the best practices and patterns for distributed development as they are mostly useful for project tracking and monitoring in the methodology.

### *E. Supportive Features*

The last proposed criteria set are about admirable features in SO methodology that are taken into consideration by methodology designers. The considered criteria are:

***1- Architecture-Based***: A successful generation of an SO system is dependent on taking into account the Service-Oriented Architecture (SOA) stack. From the stack viewpoint, SOA consists of different layers that are aligned with business goals. Services are placed as building blocks in each layer. For example the most significant layers as introduced by IBM are [28]: *operational layer*, *service component layer*, *business process layer*, *consumer layer*, *integration layer*, *QoS layer* and *Information architecture layer*.

A spirit of architecture-based development plays an essential role to facilitate system development, team management, risk mitigation and business process management even non-SO ones. Consequently, a custom instantiation from SOA stack should be made in initiating the development process and refined as the development progresses in each iteration. All development process activities and design decisions are arranged based on it.

***2- Service Agility:*** One of the main reasons to tendency to SOA investments is service agility. Today, business does not remain in a stable state. Ever-changing in nature, government rules, policies, market opportunities and threats make unanticipated changes in enterprise business processes. In order to remain in the marketing contest and be alive, an enterprise should be widely adaptable with the external environment. For this purpose, services as building blocks of business processes should be designed agile to support the business changes. Agility in services can be achieved by applying OMG's Model-Driven Architecture (MDA) [29] or Layering approaches [4,7]. This criterion evaluates the methodology from the aspect of defining prescribed principles, guidelines and techniques to reach agile services.

***3- Process Agility***: Elicitation of all requirements of a large-scale distributed system at early stages of the project is not possible. Some of the requirements are discovered or understood better during system development. Therefore, responding to new requirements and changes in business occurs frequently. Appropriate techniques or guidelines to enhance the agility of the development process in accordance to changes should be addressed. For checking this criterion, we consider the degree of relationship between SO methodology and Agile Manifesto such as flexible plan, active user involvement, short release time and increase in communication and feedback that will lead to an increase in agility of the development process to quickly respond to future changes.

***4- Maturity Level:*** Service oriented maturity models such as SOAMM or SIMM help an enterprise to achieve certain maturity levels. Although, since there is no agreement (yet) on SOA maturity models, applying a methodology should increase the level of service orientation in an enterprise. This criterion evaluates methodology activities to cover defined service orientation maturity disciplines.

***5- Tool Support***: This criterion refers to availability of appropriate tools for using in coherence with the methodology. Incorporating tools should be considered for supporting business process modeling, service modeling, SLA monitoring, and support for Enterprise Service Bus (ESB) and Middleware.

## V. Evaluation MSOAM Using Proposed Evaluation Framework

In this section, we conduct a criteria-based evaluation of the MSOAM methodology based on the proposed evaluation framework. We selected MSOAM to analyze, since its documents are fully available. As mentioned in section 3, for developing evaluation framework criteria we have used feature analysis. For a better and more understandable representation of the analysis results we used descriptive degrees as: Not Addressed, Low, Medium, and High.

Evaluating MSOAM with the proposed evaluation framework highlights its strengths and weaknesses. MSOAM process introduces a series of best practices and a formal step-by-step process for service-oriented analysis and design. It strives to be an abstract process. However, MSOAM lacks attention to define appropriate roles, activities and modeling language issues. Tables I through V show the results of evaluating the MSOAM methodology using the proposed evaluation framework.

TABLE I. RESULTS OF APPLYING THE DEVELOPMENT PROCESS CRITERIA

| Criterion | MSOAM Evaluation | Degree |
|---|---|---|
| Clarity and Consistency of Definition | Methodology has a clear definition for service analysis and design phase, but it lacks the definition for other phases for example construction | Medium |
| Coverage of the Generic Development Lifecycle Activities | Methodology defines a complete generic life cycle. It starts with service modeling and design and stops in construction phase. | Low |
| Support Umbrella Activities | There is no definition for risk management and project management | Not Addressed |
| Smooth and Seamless Transition between Phases, Stages and Activities | A step-by-step technique for service modeling is defined. There are techniques for handling other phases. | Medium |
| Basis in the Requirements (Functional and Non-Functional) | Requirements should be gathered before using methodology. No activities prescribed. | Not Addressed |
| Tangibility of Artifacts, and Traceability to Requirements | Due to lack of requirement engineering, traceability can be in risk | Low |
| Manageability of Complexity | Services are modeled and designed through a step-by-step process. But no technique is defined for architecture design. | Medium |
| Extensibility, Configurability, Flexibility and Scalability | MSOAM's development process can be run in top-down, bottom-up and agile manner. Hence, it is configurable. | Medium |
| Practicability and Practicality | Development process is practicable through providing effective activities for top-down, bottom-up and agile strategies for development. Development process is practical because it supports pragmatic techniques for service-oriented analysis and design independently from any specific tools. | High |
| Application Scope | MSOAM is a simple development process for information systems. It can be extended to fit different scopes. | Information System |
| Evolutionary or Revolutionary | Development process is defined from scratch, regardless of other development processes. | Revolutionary |
| Language or Technology | MSOAM does not target at a specific language or technology. It does not refer to the construction phase. | High |

TABLE II. RESULTS OF APPLYING THE MODELING LANGUAGE CRITERIA

| Criterion | MSOAM Evaluation | Degree |
|---|---|---|
| Support for Different Model Views | - | Not Addressed |
| Analyzability | - | Not Addressed |
| Techniques for Tackling Model Inconsistency and Managing Model Complexity | - | Not Addressed |
| Preciseness | - | Not Addressed |
| Simplicity to Learn and Use | - | Not Addressed |

TABLE III. RESULTS OF APPLYING THE SERVICE-ORIENTED ACTIVITIES CRITERIA

| Criterion | MSOAM Evaluation | Degree |
|---|---|---|
| Business Modeling | MSOAM only refers to it. Business modeling is done apart from process usually. | Medium |
| Service Oriented Analysis & Design (SOAD) | Service-oriented analysis and service-design phases are designed to support this criterion strongly. | High |
| Service Quality Attributes | - | Not Addressed |
| Service Provisioning and Consuming | - | Not Addressed |
| Service Testing | - | Not Addressed |
| Service Versioning and Evolution | - | Not Addressed |
| Adaptable with Legacy Systems | - | Not Addressed |
| Cost Estimation | - | Not Addressed |

TABLE IV. RESULTS OF APPLYING THE SERVICE-ORIENTED UMBRELLA ACTIVITIES CRITERIA

| Criterion | MSOAM Evaluation | Degree |
|---|---|---|
| Service Level Agreement (SLA) Monitoring | - | Not Addressed |
| Support of Governance | - | Not Addressed |
| People Management | - | Not Addressed |
| Distributed Software Development (DSD) Techniques | - | Not Addressed |

TABLE V. RESULTS OF APPLYING THE SUPPORTIVE FEATURES CRITERIA

| Criterion | MSOAM Evaluation | Degree |
|---|---|---|
| Architecture-Based | - | Not Addressed |
| Service Agility | MSOAM defines service layer mechanism to inspire agility to services. These three primary layers are *business process layer, service interface layer* and *application layer*. These layers enhance decoupling between granular services in layers. | High |
| Process Agility | MSOAM prescribes two strategies for having an agile process, bottom-up strategy and agile strategy. | High |
| Maturity Level | According to comparison between SOAMM with MSOMA's activities, MSOAM enhances the enterprise maturity to the level of two. | Level 2 |
| Tool Support | No tool is defined yet to support the development process. | Not Addressed |

## VI. CONCLUSION AND FUTURE WORK

Selecting a suitable methodology from the set of SO methodologies is still difficult. In this paper, we proposed a methodology evaluation framework to address detailed evaluation of different aspects of SO methodologies. This framework defines a set of criteria for identifying weaknesses and strengths, and comparing SO methodologies. Project managers can use evaluation results to select an appropriate methodology. Moreover, methodology engineers can use evaluation results as a guide for adaptation, extending, meta-modeling/instantiating of SO methodology.

Although our evaluation framework is subjective due to the use of feature analysis technique in its development, we used meta-criteria for eliminating the overlapping, redundancy and inconsistency between the criteria. One of the future works can be trying to define quantitative criteria set for the evaluation framework. Another future work can focus on refining the evaluation framework through applying it to other SO methodologies, and defining more evaluation criteria.

As the main future work, we will focus on extracting a set of process patterns from SO methodologies. Ambler defines a process pattern as "a collection of general techniques, actions, and/or tasks (activities) for developing object-oriented software" [30]. Process patterns describe commonality of process fragments in software development methodologies. Hence, our process patterns will be distillate and will show all recurring activities in different SO methodologies. They are resulted from the abstraction view on development process of SO methodologies. Process patterns provide a generic life cycle and can be instantiated for constructing a new custom SO based software development process. Process pattern is an essential part of Assembly-Based Situational Method Engineering (SME) [31]. They store reusable method fragments in method repository. Then by the SME, a new methodology is constructed by selecting and assembling process patterns from method repository according to predefined requirements of the project situation at hand. In addition, process patterns are useful for configuring, improving and comparing existing SO methodologies.